# Data-Driven Smile Design: Personalized Dental Esthetics Outcome Using Deep Learning

[1]Marcus Lin、[2]Jennifer Lai
[1]marcuslinky@gmail.com、[2]nivealai@mail2000.com.tw

**Abstract**

A bright smile is often regarded as the best form of makeup, serving not only functional roles in pronunciation and mastication but also playing a crucial part in enhancing self-confidence through aesthetic appeal. As more individuals seek aesthetic dental treatments, the challenge of balancing aesthetic desires with functional requirements becomes increasingly complex for dental practitioners. Traditional methods of smile design have relied heavily on the expertise of dentists, often utilizing plaster models and hand-drawn illustrations, which can leave patients uncertain about treatment outcomes. The integration of digital technologies, pioneered by Dr. Christian Coachman in 2007, has revolutionized this process by allowing for photographic and videographic analyses, thereby improving communication among dental professionals and patients.

Recent advancements, particularly in artificial intelligence (AI) and big data, have further transformed anterior smile aesthetics. Contemporary AI technologies possess the capability to analyze facial features and generate personalized smile design proposals. However, the output is often influenced by the implicit aesthetic biases of practitioners or the algorithm's training data, which may result in homogenized outcomes that lack true individualization. This study aims to develop a comprehensive system that integrates AI, big data, and recognition technologies to streamline the smile design process, enabling both experienced and novice dentists to produce aesthetically pleasing results efficiently.

The proposed system architecture includes a Facial Feature Extraction Module and an Image Generation Module, which together facilitate the rapid generation of smile design options. By defining user demographics and crafting user stories, the study addresses the diverse needs of dental practitioners and patients alike. The research plan outlines a systematic approach to development, testing, and analysis, ensuring that the system meets clinical demands while enhancing patient satisfaction.

Ultimately, this study recognizes the subjective nature of beauty and the potential limitations of external APIs in the Image Generation Module. Future iterations of the system may incorporate user data to refine design outputs and explore the integration of virtual and augmented reality technologies for real-time previews. Additionally, the data collected can be utilized for in-depth analyses of aesthetic preferences, contributing to a more nuanced understanding of smile design in dental practice.

## 1. Introduction

There is a saying that "a bright smile is the best makeup." The maxillary anterior teeth not only serve the functions of pronunciation and chewing, but also need to consider esthetic factors, appropriately displaying teeth while smiling and coordinating with different facial expressions to enhance our confidence in our smiles. More and more people come to dental clinics to seek aesthetic treatment for this reason. However, it's not easy to meet all requirements among aesthetics, pronunciation, and chewing function. As a dentist, we often need to try multiple attempts to achieve a result that the patient prefers.

### 1-1. Traditional Considerations in Anterior Smile Aesthetics

Traditional dental aesthetic design has historically relied on the expertise and technical skills of dental practitioners, utilizing plaster models or hand-drawn illustrations to create smile curves. This conventional approach often leaves patients unable to accurately anticipate the outcomes of their treatments. In 2007, Dr. Christian Coachman and his team pioneered the integration of digital technology with smile design, employing photographic and videographic analyses of patients' facial features through advanced computer software. This innovation not only facilitates interdisciplinary collaboration among dental professionals but also enhances communication between dentists and technicians, allowing both parties, as well as patients, to preview and discuss the anticipated final results prior to treatment[1].

Subsequent advancements in technology, particularly breakthroughs in intraoral and facial scanning techniques, have been further augmented by the integration of computed tomography imaging. These developments enable the rapid production of 3D-printed prosthetics using CAD/CAM technology following the design phase, thereby allowing patients to experience the final outcomes of their smile treatments in advance [2]. Following 2020, the emergence of artificial intelligence (AI) technologies has led to the development of various applications capable of automatically analyzing patients' facial and smile characteristics, swiftly generating optimal smile design proposals. However, it is important to note that these designs predominantly rely on the subjective judgments of dental professionals and technicians, often resulting in standardized outputs that yield anterior designs lacking in individuality and esthetic nuance.

## 1-2. Current Considerations in Anterior Smile Aesthetics

The design of anterior smile aesthetics currently encompasses several critical factors:

1. Shape and Size of Teeth: The shape and size of the teeth should align with the patient's facial features to ensure a natural and harmonious appearance. [4,5]
2. Color of Teeth: The color of the teeth must be consistent with the patient's skin tone to enhance overall aesthetic appeal.
3. Smile Curve: An aesthetically pleasing smile curve aligns harmoniously with the contour and placement of the lips, contributing to an overall more appealing smile.
4. Arrangement of Teeth: The teeth should be arranged in a proper position and symmetrical manner [3].

For young or less experienced dental practitioners, the lack of clinical experience may hinder their ability to design an appearance that meets both the aesthetic and practical needs of patients. This study aims to leverage artificial intelligence (AI) technology to assist in generating multiple anterior design options based on dental aesthetics, thereby providing a rapid solution to such challenges.

For seasoned dentists, although they possess extensive experience, the varying focuses of continuing education post-graduation may lead to individual design preferences. Therefore, AI can offer a diverse array of design options, enabling these practitioners to propose several alternatives that are more likely to satisfy their patients in a shorter timeframe.

## 2. Literature Review

According to research by Vig and Brundo (1978) published in the Journal of Prosthetic Dentistry, the amount of maxillary incisors exposed when the lip muscles are relaxed tends to decrease with advanced age [6]. However, not all individuals are satisfied with their smiles. We find out that even middle-aged and elderly patients express a desire for 3-4 mm of incisor exposure during lip relaxation when undergoing reconstruction of their anterior teeth. This highlights a growing trend in recent years where patients are increasingly vocal about their preferences and opinions regarding anterior tooth aesthetics. Consequently, utilizing big data to provide patients with visual options for selection has emerged as a significant trend in dental practice.

In 2023, Ali and colleagues developed a deep learning-based system specifically designed for digital smile design. This system employs convolutional neural networks (CNNs) to analyze and generate smile designs. The study utilized a substantial dataset of facial and dental images to train the deep learning model, ensuring its capability to identify and produce aesthetically pleasing designs accurately [7].

Furthermore, Ceylan and colleagues (2023) employed standardized assessment tools to compare the aesthetic outcomes of smile designs generated by AI with those created by dental professionals. Through questionnaires evaluating aesthetics, functionality, and patient satisfaction, the results demonstrated that AI-generated smile designs were comparable to those produced by dental professionals in certain aesthetic parameters, with both groups achieving satisfactory levels of patient contentment. The findings of this study underscore the potential of artificial intelligence to enhance precision and objectivity in aesthetic design processes within the field of dentistry [8].

Mohsin et al. (2025) developed and clinically validated an AI-based smile design software that integrates convolutional neural networks (CNNs) and generative adversarial networks (GANs). The system demonstrated significant advantages over traditional methods in terms of aesthetic outcomes, patient satisfaction, and design efficiency. This study not only provides empirical evidence for the application of CNN-GAN integration in smile design but also highlights the practical potential of artificial intelligence in addressing issues arising from the overreliance on clinicians' subjective judgment. However, the system proposed by Mohsin et al. primarily focuses on the quantitative evaluation of clinical manifestations, with relatively limited attention to the automation of internal processes and the flexibility of the overall system architecture. In our study, the system focuses on integrating a third-party aesthetic evaluation mechanism (Face++) to automatically select the most aesthetically pleasing design outcomes by setting aesthetic thresholds and implementing an iterative screening process, thereby effectively reducing reliance on manual intervention and subjective judgment. In terms of system architecture, a modular design is adopted, offering high flexibility and scalability. This allows for the seamless integration of various recognition models and generative algorithms in the future, and further extension to application scenarios such as virtual reality previews, multimodal inputs, or integration with clinical data. More importantly, the system enables patients to directly select their preferred smile designs from automatically generated and aesthetically filtered options, thereby enhancing patient engagement and improving communication efficiency between clinicians and patients [9].

## 3. Methodology

The objective of this study is to develop a comprehensive system that integrates artificial intelligence (AI), big data, and recognition technologies to facilitate the generation of smile design outcomes. This system aims to assist dental practitioners in executing smile designs more efficiently while enabling less experienced dentists to produce natural and aesthetically pleasing results.

The proposed system architecture is illustrated in Fig. 1, and the following sections provide a detailed explanation of each module:

A. Facial Feature Extraction Module:

The process begins with the dentist capturing standardized high-resolution facial photographs of the patient. This module utilizes MediaPipe Face Mesh to extract 468 facial landmarks, capturing precise geometric information related to facial proportions, smile curvature, and overall facial contour. Since MediaPipe itself does not provide face shape classification functionality, the extracted facial landmarks serve as input features for a separate, custom-trained face shape classification model that performs the actual classification. To train the initial facial classification model, this study adopts the SCUT-FBP5500 dataset, which includes 5,500 facial images annotated with various face shapes. From this dataset, approximately 500 images that meet the specific requirements for smile design (e.g., frontal views with clear expressions) will be manually selected. Each selected image will undergo the following six data augmentation operations to increase diversity and volume:
1. the original image.
2. increased brightness and reduced contrast.
3. reduced brightness and increased contrast.
4. horizontal flipping of the original image.
5. horizontal flipping of item 2.
6. horizontal flipping of item 3.

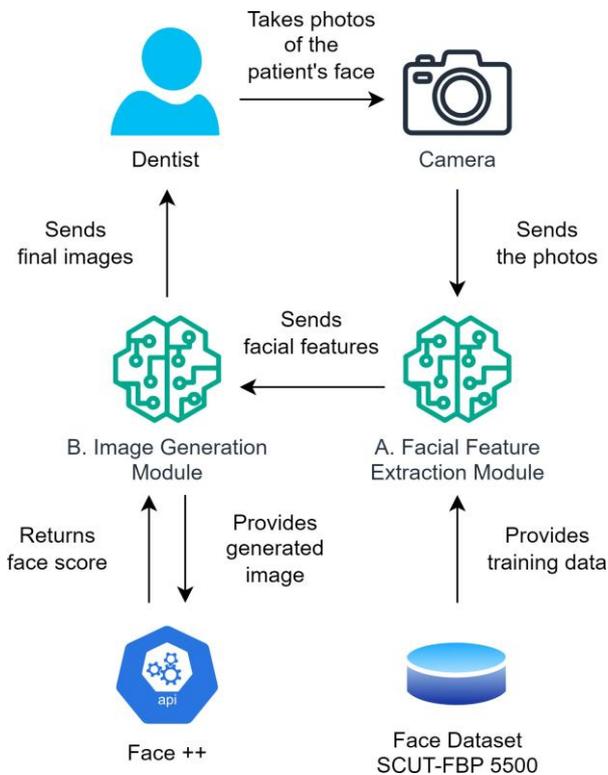

Fig. 1 System Architecture Diagram

These augmentations will result in a total of approximately 3,000 training samples. To ensure model robustness and mitigate overfitting, a fivefold cross-validation strategy will be applied during training.

Following the release of the first system version, users will be invited to consent to the inclusion of their anonymized clinical photographs for future model improvement. With patient approval, these real-world images will be incorporated into the training dataset to enhance recognition accuracy. Once a sufficient volume of high-quality internal data is collected, the original 3,000 augmented samples will be gradually phased out and replaced, enabling the development of a second, more application-specific model version.

This approach combines the accuracy of a pre-trained landmark detector with the flexibility of a supervised learning classifier to provide individualized and precise facial feature analysis.

B. Image Generation Module:

This module receives the facial features and face shape classification results extracted from Module A and utilizes generative adversarial network technology (StyleGAN2) to produce multiple realistic smiling images. The process begins by encoding the patient's facial image into a numerical representation suitable for machine interpretation. Subsequently, InterfaceGAN is employed to modify smile-related attributes, producing diverse variations in smile expressions.

Each generated image is submitted to Face++, an AI platform offering an aesthetic evaluation API. Face++ assigns an attractiveness score based on machine learning models trained on large-scale facial datasets. Only images with scores exceeding a predefined aesthetic threshold (e.g., 70/100) are retained. An iterative refinement loop is employed to generate smile designs, and the process proceeds until five outputs meeting the aesthetic threshold are selected. These selected designs are compiled into a collection and returned to the dentist for patient selection and treatment planning. The system architecture integrates digital imaging, facial feature analysis, and artificial intelligence to facilitate a data-driven smile design workflow for dental professionals.

With the facial feature extraction module and the image generation module described above, we are able to establish a data-driven, patient-centered smile design workflow. The system combines precise biometric technology with objective AI aesthetic scoring, ensuring consistent aesthetic standards and facilitating effective communication between dentists and patients, thereby reducing the risk of disputes arising from differing aesthetic expectations. Meanwhile, the system greatly simplifies the dentist's workflow, requiring only that dentists capture and upload standardized facial photographs, then await processing and design generation. Finally, dentists can make informed decisions based on the generated outcomes. A detailed overview of this workflow is illustrated in Fig. 2.

To initiate a smile design plan, the dentist captures high-resolution photographs of the patient's face, which are then uploaded to the Facial Feature Extraction Module. This module extracts key facial parameters for further processing. The extracted data is transmitted to the Image Generation Module, which creates multiple simulated smile designs. Each design is evaluated by the AI-based platform Face++, and only those scoring 70 or above are retained as candidate images. These selected images are presented to the patient for review and preference selection. The dentist then uses the chosen design as a reference to develop the final treatment plan, integrating it into the clinical workflow.

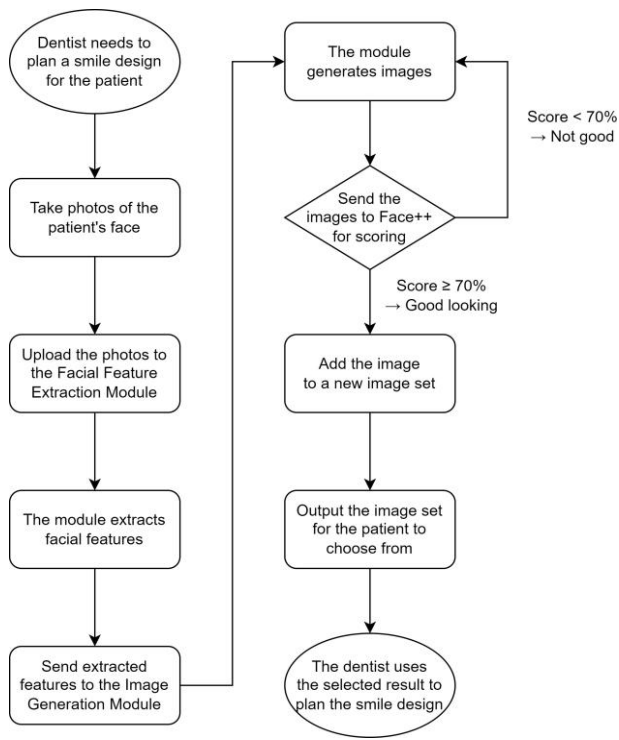

Fig. 2 System Flowchart

## 4. Research Plan

This study plans to adopt the following six steps for system development, with each step detailing the associated tasks and expected outcomes:

1. Defining the User Base

By clearly identifying the target demographic, we can more accurately integrate AI technologies to assist in dental clinical decision-making, thereby meeting user needs and enhancing the overall user experience. The primary users of this system will be dental practitioners, particularly those involved in anterior aesthetics and full-mouth rehabilitation. Additionally, patients who place a high value on smile curvature or facial harmony represent a potential user group. For these individuals, the ability to preview and customize their smile in real-time can significantly improve communication between the dentist and the patient, as well as enhance the patient's confidence in the treatment process. Activities during this phase may include needs interviews and clinical case analyses, with outputs potentially encompassing usage scenarios and specifications documents.

2. Designing User Stories

To accurately address the needs of the defined user groups, user stories will be crafted based on various usage scenarios that simulate real-life situations while validating the system's practicality. Three distinct scenarios are proposed:
Type A: Highly aesthetically demanding patients
Scenario: Patients desire to enhance their appearance but struggle to articulate their desired outcomes, leading to results that may not align with their expectations.
Expectation: Patients hope to have multiple design options to identify their preferred aesthetic.

Type B: Busy specialists
Scenario: With a busy and highly compact schedule each day, dedicating time to personally design smiles is time-consuming, leaving little room for breaks or patient coordination.
Expectation: The ability to quickly generate a large number of smile design outcomes.

Type C: Novice dentists
Scenario: Due to a lack of experience, novice dentists frequently seek guidance from more seasoned colleagues on how to make adjustments.
Expectation: A desire for quicker access to accurate feedback.

3. System analysis (SA) and system design (SD)

This phase will translate the aforementioned user requirements into actionable technical solutions, ensuring that the system can efficiently and accurately address these challenges. This study aims to integrate AI models to establish an automated system for generating smile design outcomes. Tasks may include confirming the architecture, selecting technologies, and defining modules, all referencing the system architecture diagram (Fig. 1) and the system flowchart (Fig. 2). The final outputs will serve as the foundation for subsequent system development.

4. System Development
System development will include the necessary model training and the implementation of the two core modules:
A. Facial Feature Extraction Module.
B. Image Generation Module.
These tasks will be carried out based on the methodologies and architecture introduced in Chapter 3.

5. Testing
This study is scheduled to conduct two phases of testing:
(1) Internal Testing: Developers will ensure that all target functionalities operate correctly and without anomalies.
(2) External Testing: In collaboration with Anident, we will recruit dentists as test subjects, aiming for a sample size of 100 participants over a six-month experimental period. During this time, dentists will be instructed to utilize the system for clinical smile design applications. Upon completion of the experiment, participants will fill out usability and ease-of-use feedback questionnaires.

6. Analyzing research outcomes and issues
Based on the feedback received, we will explore whether AI applications in smile design are beneficial, ultimately documenting any unresolved issues or areas requiring improvement.

## 5. Discussion

Given the inherently subjective nature of facial aesthetics, it is often challenging to fulfill every patient's individual expectations for smile design outcomes. To address this limitation, it becomes critical to incorporate objective evaluative frameworks—such as leveraging aggregated public aesthetic assessments—to reduce the disparity between patient desires and clinically achievable results. By integrating such data-driven insights, dental professionals can better align their treatment planning with socially validated standards of beauty. However, it is important to acknowledge that the **Image Generation Module** in this system is dependent on third-party APIs, such as Face++, for aesthetic scoring. This reliance introduces potential vulnerabilities, including service interruptions, changes in access policies, or additional costs imposed by the external provider, which may affect the scalability and sustainability of the overall system.

In addition, the iterations of the system could incorporate the collection of users' before-and-after data, integrating this information into a new dataset that can be utilized alongside the original model. There are also plans to enhance the system with virtual reality (VR) and augmented reality (AR) technologies, allowing patients to preview changes in their smile design in time.

Moreover, the outcomes and data generated by this system can be further integrated into treatment plans. Based on the aesthetic preferences selected by the patient, the system could suggest necessary orthodontic adjustments, soft tissue and hard tissue augmentations, and prosthetic restorations, thereby estimating the required treatment duration and associated costs for the patient's consideration. Additionally, through interactions between AI and the dentist, the system could predict achievable surgical outcomes and recommend customized design options.

Finally, the data collected can be employed in data mining to conduct in-depth analyses of the correlations between various smile designs and aesthetic appeal. For instance, it may be found that a patient with a round face paired with oval-shaped teeth receives higher aesthetic ratings; or with a big nose paired with big central incisors receives higher aesthetic ratings, etc.